\documentclass{article}

\usepackage{arxiv}

\usepackage[utf8]{inputenc} 
\usepackage[T1]{fontenc}    
\usepackage{hyperref}       
\usepackage{url}            
\usepackage{booktabs}       
\usepackage{amsfonts}       
\usepackage{nicefrac}       
\usepackage{microtype}      
\usepackage{lipsum}

\usepackage{lineno}
\usepackage{amsmath}
\usepackage{subfig}
\usepackage{graphicx}
\usepackage{algorithm}
\usepackage{algpseudocode}

\title{On sampling of scattering phase functions}
\author{
   Jianing Zhang\thanks{Use footnote for providing further
    information about author (webpage, alternative
    address)---\emph{not} for acknowledging funding agencies.} \\
  Dalian University of Technology\\
  Panjin, 124221, China \\
  \texttt{fugiya@dlut.edu.cn} \\
}

\begin{document}
\maketitle

\begin{abstract}
Monte Carlo radiative transfer, which has been demonstrated as a successful algorithm for modelling radiation transport through the astrophysical medium, relies on sampling of scattering phase functions. We review several classic sampling algorithms such as the tabulated method and the accept–reject method for sampling the scattering phase function. The tabulated method uses a piecewise constant approximation for the true scattering phase function; we improve its sampling performance on a small scattering angle by using piecewise linear and piecewise log-linear approximations. It has previously been believed that certain complicated analytic phase functions such as the Fournier-Forand phase function cannot be simulated without approximations. We show that the tabulated method combined with the accept-reject method can be applied to sample such complicated scattering phase functions accurately. Furthermore, we introduce the Gibbs sampling method for sampling complicated approximate analytic phase functions. In addition, we propose a new modified Henyey-Greenstein phase function with exponential decay terms for modelling realistic dust scattering. Based on Monte Carlo simulations of radiative transfer through a plane-parallel medium, we also demonstrate that the result simulated with the new phase function can provide a good fit to the result simulated with the realistic dust phase function. 
\end{abstract}


\section{Introduction}

Monte Carlo radiative transfer (MCRT) \cite{chandrasekhar, whitney, steinacker} is a widely used numerical algorithm for studying transport of radiation in interstellar mediums \cite{draine}, and the planetary atmosphere \cite{manfred}.  The MCRT primarily relies on producing pseudo random numbers to evaluate the integral form of the radiative transfer equation with statistical simulation. Considerable efforts have been devoted to produce sequences of random numbers with high qualities in order to implement a Monte Carlo simulation. In MCRT, a key step is to sample scattering directions from the phase function, which plays a fundamental role in understanding the physical properties of the medium \cite{wiscombe, Mobley:02, Freda:07,  Piskozub:11,  Gkioulekas:2013, Pitarch:16, ZHANG2016325, Marinyuk:17, Tuchow:16}. ]. Hence, sampling of scattering phase functions is related to the accuracy of the simulated results in MCRT, and consequently, efficient sampling algorithms are particularly essential and in high demand.

For an approximate analytic phase function, it is effective to conduct simulations using the direct method \cite{gentle, whitney} by performing an inverse transform of the cumulative distribution function (CDF) corresponding to the scattering phase function. For realistic phase functions such as the Mie's phase function, an analytic form of their inverse CDF may not exist or computing their values may be too expensive. Therefore,  sampling has to be conducted in different ways. For example, the tabulated method \cite{Toublanc:96, Zijp:94} is often used for sampling such complicated phase functions by constructing a look-up table, where random numbers are obtained by linear interpolation. As an approximate sampling method, the performance of the tabulated method is dependent on the number of interpolation points. However, to sample realistic phase functions (with peaks near the forward and backward directions) accurately, a considerably greater number of points are required to obtain a reliable result \cite{Naglic:17}. Poor sampling near forward and backward directions has a large negative impact on the simulation accuracy of radiation transport \cite{Borovoi:13,  Naglic:17}. The accept-reject method \cite{gentle, pharr} is another available method. For the Rayleigh phase function, Frisvad \cite{Frisvad:11} reviewed several sampling methods and concluded that the simple accept-reject method was the most efficient. For realistic phase functions \cite{Mobley:02, Freda:07,  Piskozub:11, Gkioulekas:2013, Marinyuk:17} with a highly forward scattering peak, the simple accept-reject method poses certain limitations owing to the high rejection rate \cite{luc}.

In this paper, we propose several improvements for the widely used simple tabulated method for sampling realistic phase functions. Although some of these improvements may have been previously designed for general purpose \cite{luc} and may be known to some researchers in the radiative transfer community, there have been no detailed descriptions on them in the literature, particularly on how to correctly sample scattering phase functions with these methods. Moreover, several complicated approximate phase functions such as the Fournier-Forand (FF) phase function \cite{Fournier:94, Fournier:99} have often been considered to be impossible to sample accurately. In this work, we show that these phase functions could be accurately simulated with the proposed hybrid sampling method. In addition, Gibbs sampling \cite{gentle}, a Markov chain sampling method, is also suggested for generating random numbers for sampling complicated approximate analytic phase functions. As examples, detailed Gibbs sampling procedures are described for the Draine phase function \cite{Draine:03} and a new modified Henyey-Greenstein phase function with two exponential decay terms. This new phase function is also validated based on Monte Carlo simulations. This work is expected to contribute to the fields of astronomy and earth sciences by providing detailed descriptions on how to sample difficult scattering phase functions in applications of Monte Carlo radiative transfer.

\section{Phase function}
Light scattering by small particles \cite{Bohren} is characterised by the scattering cross section and the scattering phase function. The scattering phase function is defined as the normalised differential scattering cross section \cite{Bohren}:
\begin{equation}
p(\Omega, \Omega') = \frac{1}{\sigma_{sc} (\Omega)}\frac{d\sigma_{sc} (\Omega)}{d\Omega'}
\end{equation}
where $\Omega$ and $\Omega'$ are the incident and scattered directions of light. Under this definition, the scattering phase function $p(\Omega, \Omega')$ describes the angular distribution of scattered light for a specific wavelength. If the medium is assumed to be isotropic, the scattering phase function is a function of the angle $\theta$ between $\Omega$ and $\Omega'$ only, i.e., $\cos\theta$.

The scattering phase function for small particles can be obtained numerically by solving Maxwell's equations. Various numerical methods \cite{Bohren} have been proposed to address the scattering problem. However, many approximate analytic phase functions are still widely employed owing to their simplicity and usefulness. One such phase function proposed by Henyey-Greenstein n\cite{Henyey1941} (HG) to model scattering of dust can be expressed as 
\begin{equation}
p_{HG}(\theta | g) = \frac{1}{2}\frac{1-g^2}{(1 + g^2 - 2 g \cos\theta)^{3/2}}
\end{equation}
where $-1\leq g\leq 1$. This single parameter phase function has been used to simulate the scattering properties of various particles. The parameter $g$ can be determined by the first moment $g = <\cos\theta>$. Although the HG phase function is quite successful, it fails to reproduce observations. Various numerical and experimental studies have shown that the HG phase function underestimates the possibility of small and large angle scattering. To address this problem, other scattering phase functions \cite{Bevilacqua:99, Pfeiffer:08, Vaudelle:17} were established such as the one proposed by Draine: 
\begin{eqnarray}
 p_{D}(\theta | g, \alpha) &=&  \frac{1-g^2}{2(1+g^2-2g\cos\theta)^{3/2}} \frac{1+\alpha\cos^2 \theta}{1+\alpha (1+2g^2)/3}
\end{eqnarray}
with the normalization constant $C = \frac{1}{1+\alpha(1+2g^2)/3}$. The Draine phase function tends to the Rayleigh phase function at long wavelengths and reduces to the phase function proposed by Cornette \& Shanks\cite{Cornette:92} when $\alpha = 1$.

A more sophisticated approach for modelling realistic phase functions can be established by assuming a particle size distribution and computing the ensemble average. For example, the power-law distribution for atmospheric and marine particles \cite{Fournier:99} can be expressed as: 
\begin{equation}
n(r)  = A r^{-\alpha},  \qquad  r \geq r_{\min}
\end{equation}
where $\alpha $ is a parameter, $r$ is the radius of the particle volume-equivalent sphere,  $r_{\min} ( > 0)$ is the lower bound of $r$, and $ A = (\alpha-1)r_{\min}^{\alpha - 1}$ is the normalisation constant. It should be noted that $r_{\min}$ is necessary to ensure that the distribution is reasonable, which is often ignored in the literature. Then, the ensemble average scattering phase function can be obtained by calculating the ratio of ensemble average differential scattering cross section to the ensemble average scattering cross section:
\begin{equation}
p(\theta) = \frac{\int_{r_{\min}}^{\infty} n(r) \sigma_{sc}(r) p(\theta,r) dr}{\int_{r_{\min}}^{\infty} n(r) \sigma_{sc}(r)dr}
\end{equation}
Based on the power-law particle size distribution, Fournier and Forand presented a two-parameter approximate analytic phase function  \cite{Fournier:99} where each particle scatters light with an anomalous diffraction approximation. The latest form of the FF phase function can be written as
\begin{eqnarray}
 p_{FF}(\theta) &=& \frac{1}{2 (1 -\delta )^2 \delta^{\nu}}[\nu(1-\delta) - (1-\delta^{\nu})\nonumber \\
 &+& [\delta(1-\delta^{\nu}) - \nu (1-\delta)]\sin^{-2}(\frac{\theta}{2}) ] \nonumber\\
 &+& \frac{1-\delta^{\nu}_{\pi}}{8(\delta_{\pi} - 1)\delta^{\nu}_{\pi}}(3\cos^2\theta - 1)
\end{eqnarray}
where $\nu =  \frac{3-\alpha}{2}$ and $\delta = \frac{4}{3(m-1)^2}\sin^2(\theta/2)$, $\alpha$ is the exponent of the power-law distribution, $m$ is the real part of the relative refractive index of the particle. The FF phase function can successfully model a marine environment. However, the value of $p_{FF}(\theta)$ at $\theta = 0^{\circ}$ can not be defined, which is rather peculiar. Hence, we suggest a regularised FF phase function:
\begin{eqnarray}
p'_{FF}(\theta) &=& \mathbb{P}(\theta \leq \theta_0)\mathbb{I}(\theta \leq \theta_0) \frac{1}{1-\cos\theta_0} \nonumber\\
&+&  \mathbb{P}(\theta > \theta_0) \mathbb{I}(\theta > \theta_0) p_{FF}(\theta)
\end{eqnarray}
where  $\mathbb{P} (\theta \leq \theta_0)$ represents the probability that the scattering angle is smaller than or equal to $\theta_0$ and $ \mathbb{I}(x \in S)$($S$ is a set.) is the indicator function, which is 1 if $x\in S $ and 0 otherwise .

\section{Sampling Methods}
\subsection{Inverse CDF method}
Monte Carlo simulations are based on the generalisation of random samples that are distributed according to a certain distribution. Samples from nonuniform distributions can be generated by transforming sequences of uniform random numbers on $(0, 1)$. If an inverse CDF exists, a continuous random variable $\mu$ can be generated by taking the transform of a uniform random variable $\xi$ on $(0, 1)$,
\begin{equation}
\mu = F^{-1}(\xi), \text{ and  } F(\mu) = \int^{\mu}_{-1}p(\mu')d\mu'
\end{equation}
where $\mu$ is a random variable with the desired distribution, $p(\mu)$ denotes the probability density (in this setting, the scattering phase function) and $F(\mu)$ denotes the CDF. For example, the HG phase function can be sampled easily with the inverse CDF technique:
\begin{eqnarray}
\mu &=& \frac{1}{2g}(1+g^2 - (\frac{1-g^2}{1- g + 2g\xi_1})^2)\\
\phi &=& 2\pi \xi_2
\end{eqnarray}
where $\mu = \cos\theta$, and $\xi_1$ and $\xi_2$ are random numbers uniformly distributed on $(0, 1)$.

\subsection{Accept-reject method}
It is sometimes difficult or even impossible to obtain an analytical form of the inverse CDF for certain phase functions. In such cases, the accept-reject method can be used, in which simulate the desired density $f_X$ (scattering phase function) is simulated using the similar but simpler density function $f_Y$.  

In the accept-reject method, we first sample a random number $x$ from density $f_Y$, then draw a uniform random number $\xi \sim \mathcal{U}(0,1)$, and accept $x$ if $\xi < f_X(x)/(M f_Y(x))$ ( $M$ can be chosen as the maximum of $f_X / f_Y$.). Otherwise, we return to the first step. Iterations of this  procedure will produce a sequence of random numbers from the density $f_X$. The average number of iterations for generating a random variate is proportional to $M$ ($M$ is larger than 1.). In general, if $f_X(x)$ can be written as $f_X(x) = Cg(x) f_Y(x)$, where $f_Y(x)$ is another density function and $C$ is a positive constant that makes $g(x) \in [0, 1]$,  random numbers from $ f_X(x)$ can be generated by the following procedure:
\begin{algorithm}
\begin{algorithmic}[1]
\Procedure{Accept-Reject Sampling}{}
\State \textbf{do}
\State \qquad sample $x \sim f_Y(x)$
\State \qquad sample $\xi \sim \mathcal{U}(0, 1)$   
\State \textbf{while} ($\xi > g(x)$)
\State \textbf{return} $x$
\EndProcedure
\end{algorithmic}
\end{algorithm}

\subsection{Tabulated method}
It is often difficult to simulate common phase functions for realistic particles using the simple accept–reject method owing to the highly forward scattering peak. The tabulated method is usually applied for such cases where a look-up table is constructed and then random numbers are generated using linear interpolation. In the following section, we present several tabulated methods for accurately sampling random scattering directions in Monte Carlo radiative transfer.

\begin{figure}
\centering
\includegraphics[width=0.8\linewidth]{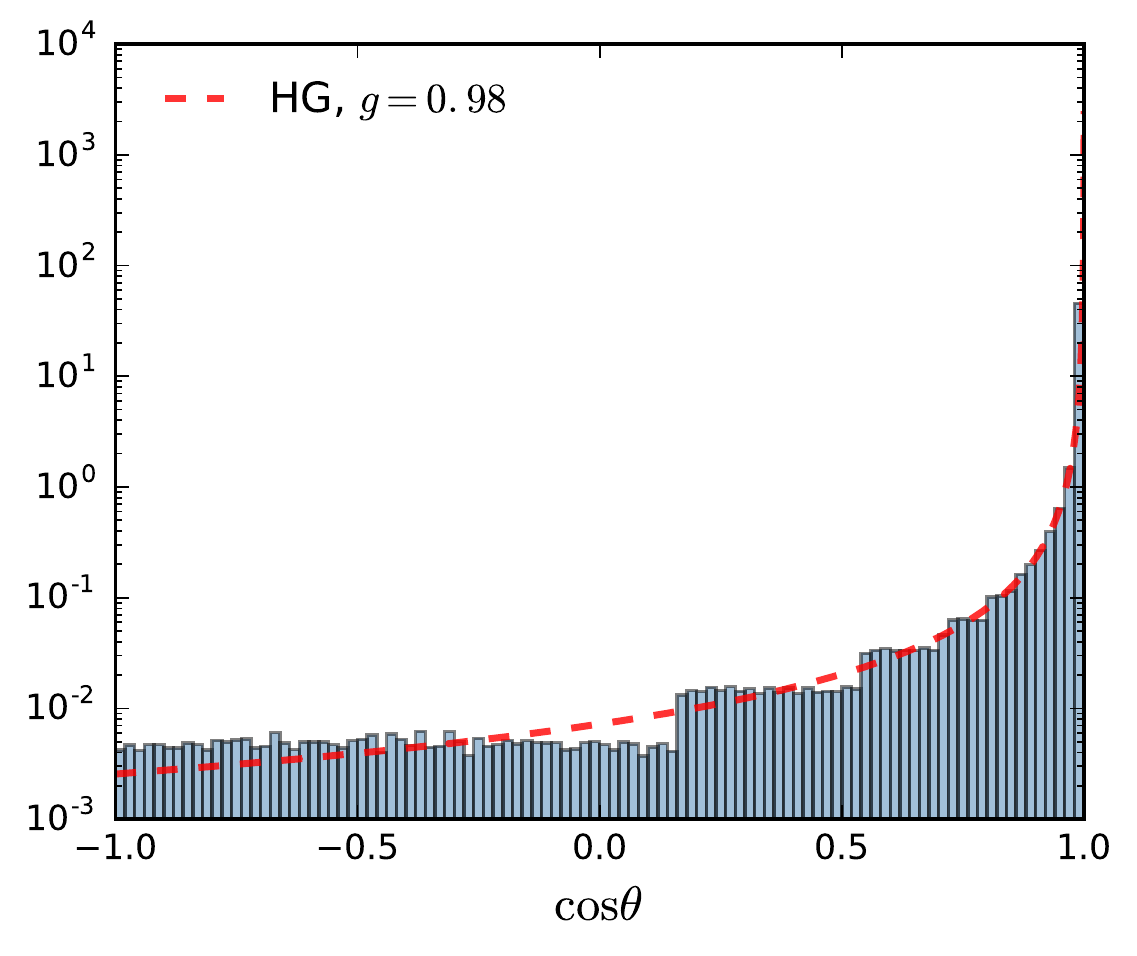}
\caption{HG phase function (red dashed line) compared with the normalised histogram of 1000000 samples generated using the tabulated method with PCA.}
\label{fig:hgpca}
\end{figure}

\begin{figure}
\centering
\includegraphics[width=0.8\linewidth]{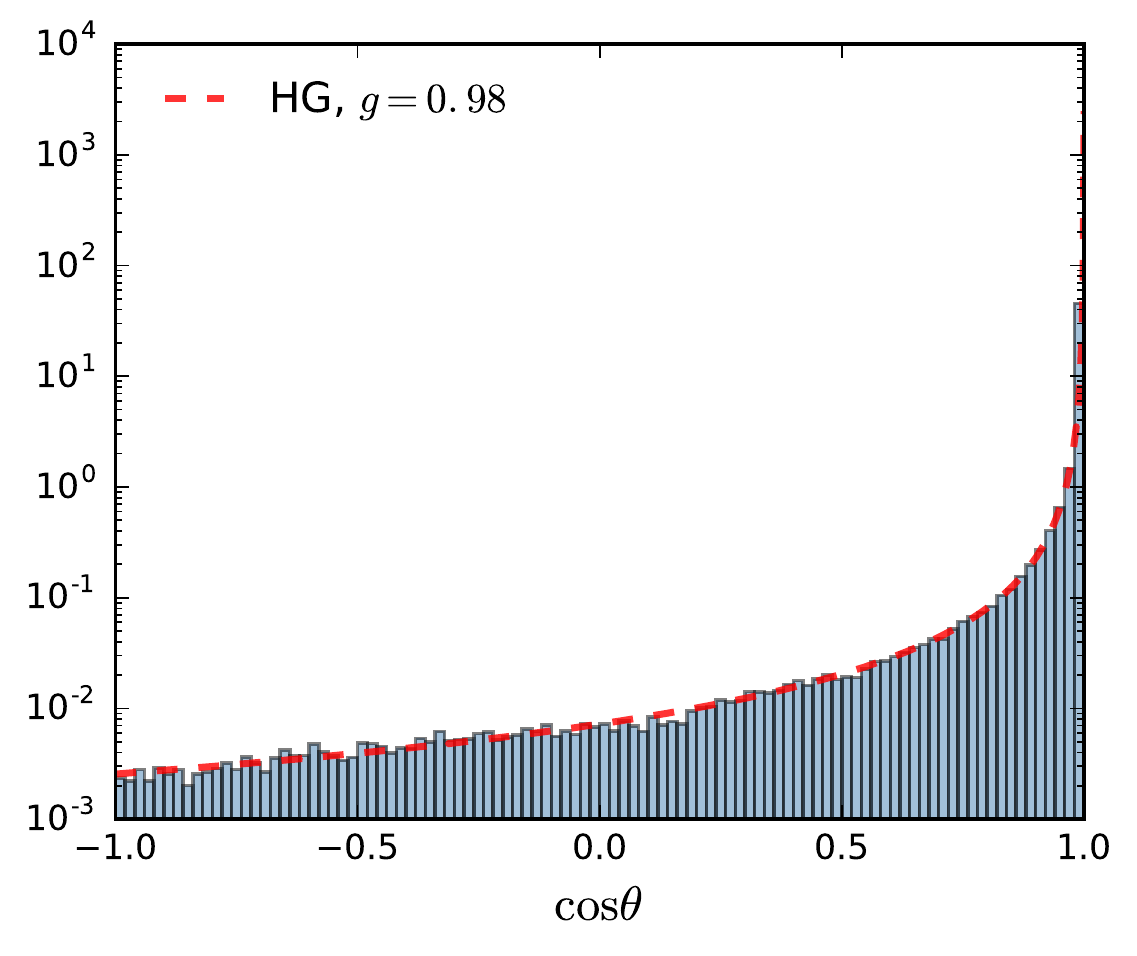}
\caption{HG phase function (red dashed line) compared with the normalised histogram of 1000000 samples generated using the hybrid tabulated method with PLA. }
\label{fig:hgpla}
\end{figure}

\begin{figure}
\centering
\includegraphics[width=0.8\linewidth]{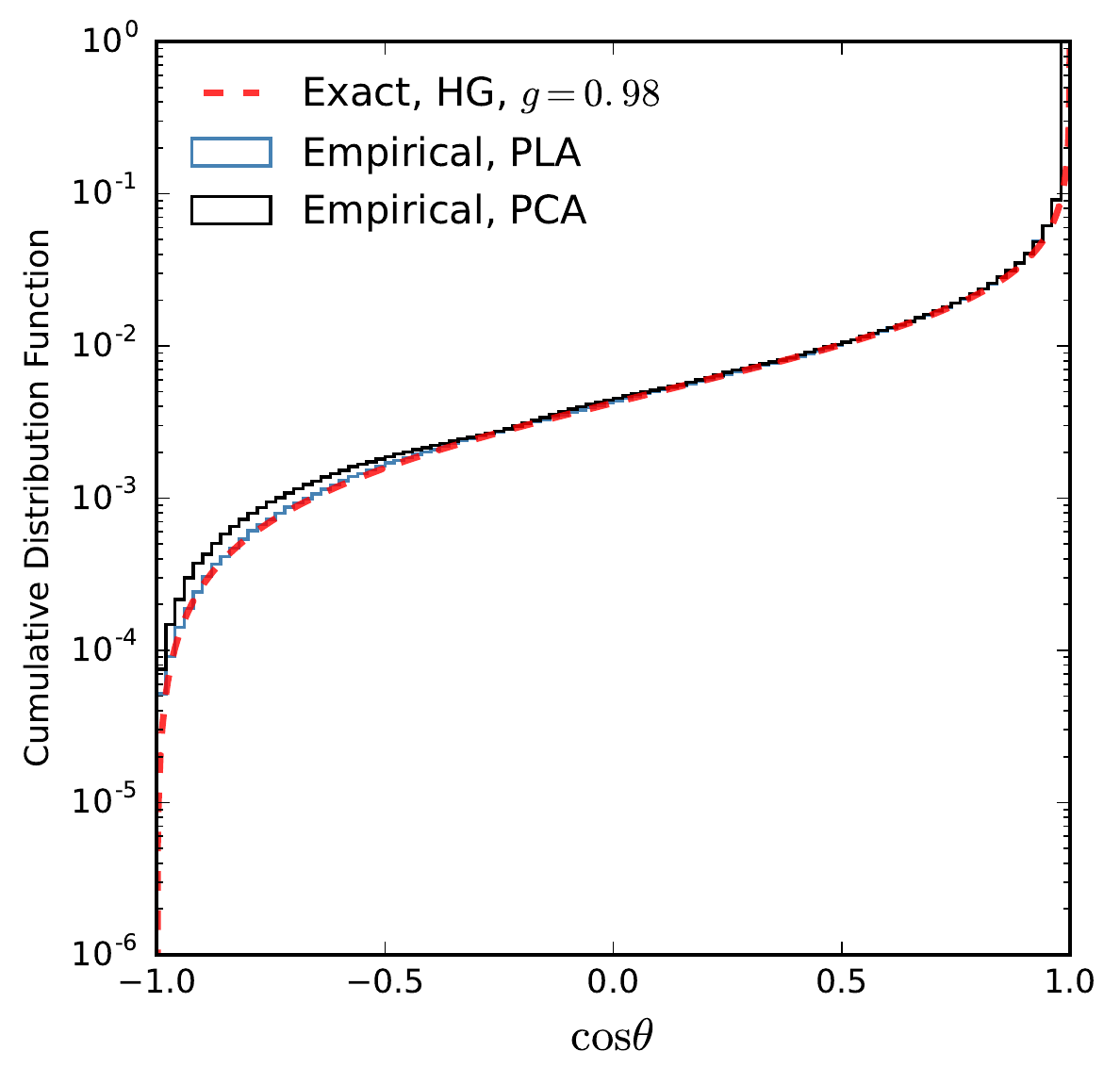}
\caption{Exact CDF of HG scattering (red dashed line) compared with the empirical CDF of HG scattering with PLA (blue stepped solid line) and PCA (black stepped solid line).}
\label{fig:hgcdf}
\end{figure}

\begin{figure}
\centering
\includegraphics[width=0.8\linewidth]{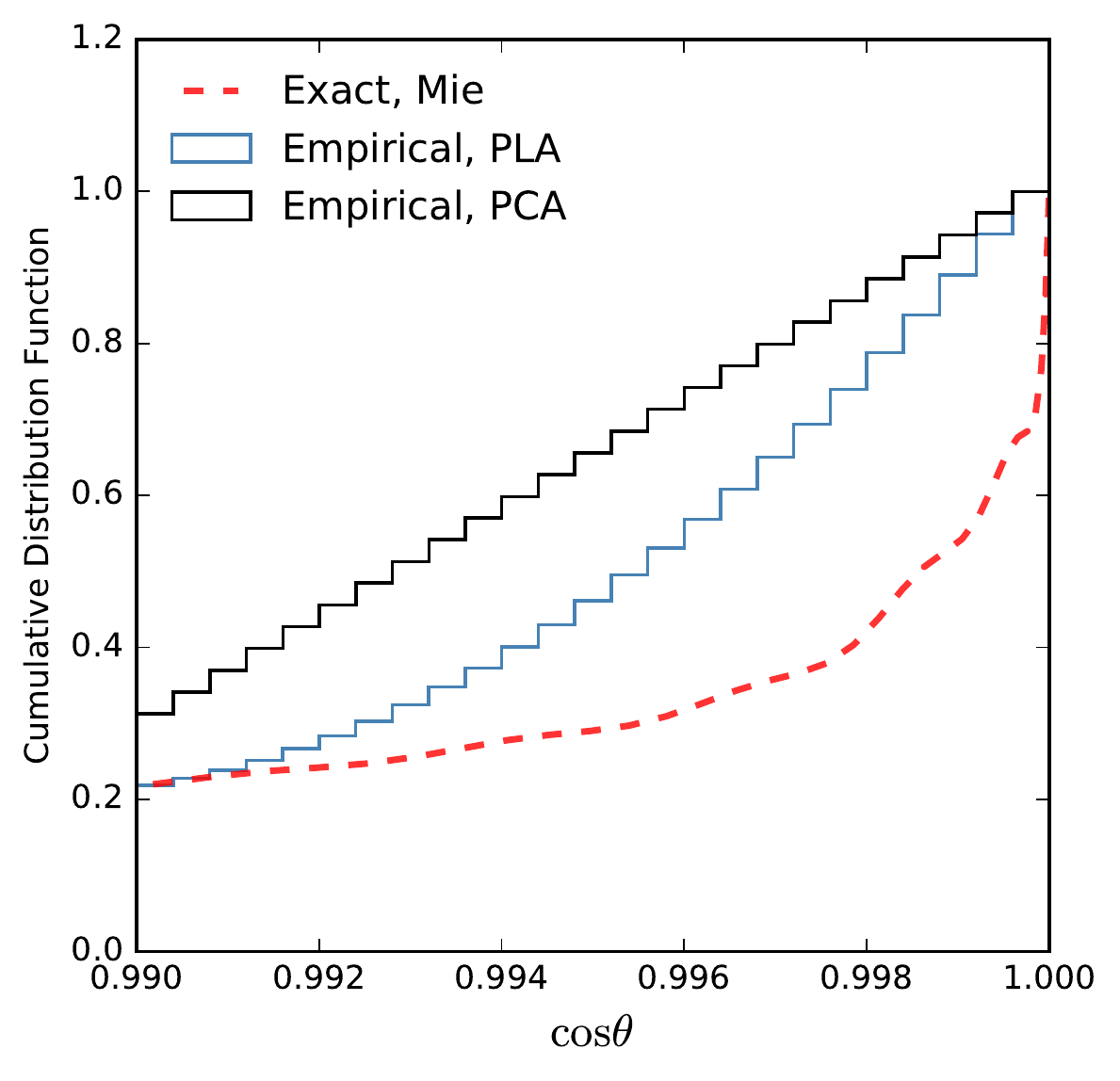}
\caption{CDF of Mie scattering (red dashed line) compared with the empirical CDF of Mie scattering with PLA (blue stepped solid line) and PCA (black stepped solid line). Mie phase function was calculated for a spherical particle with a radius of $2\mu m$, a refractive index of 1.42, buried in a medium with a refractive index of 1.352, illuminated by incident light with a wavelength of $600 nm$.}
\label{fig:miecdf}
\end{figure}

\begin{figure}[ht]
\centering
\subfloat[]{\includegraphics[width=0.6\linewidth]{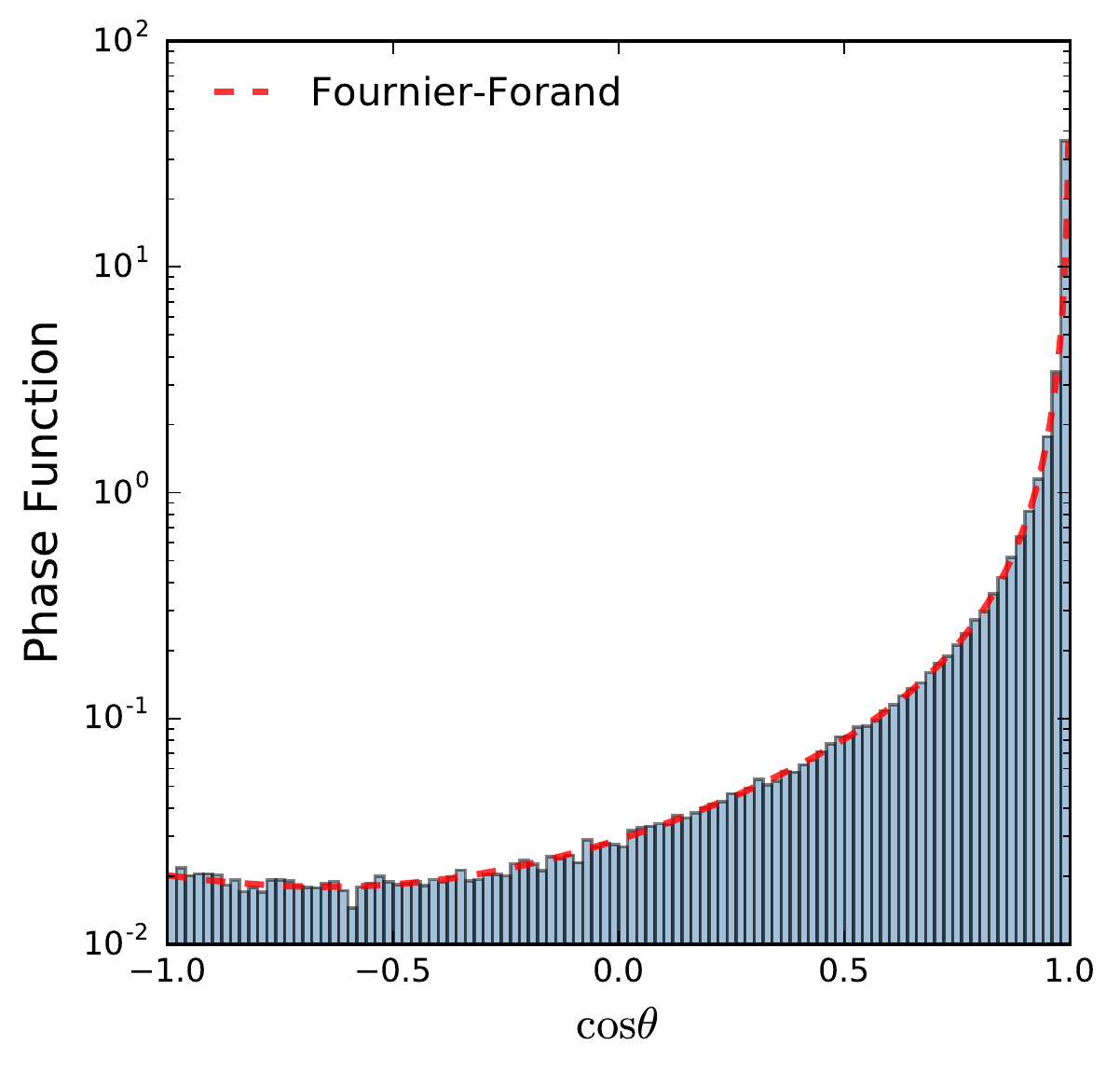}\label{ff:a}}\\
\centering
\subfloat[]{\includegraphics[width=0.6\linewidth]{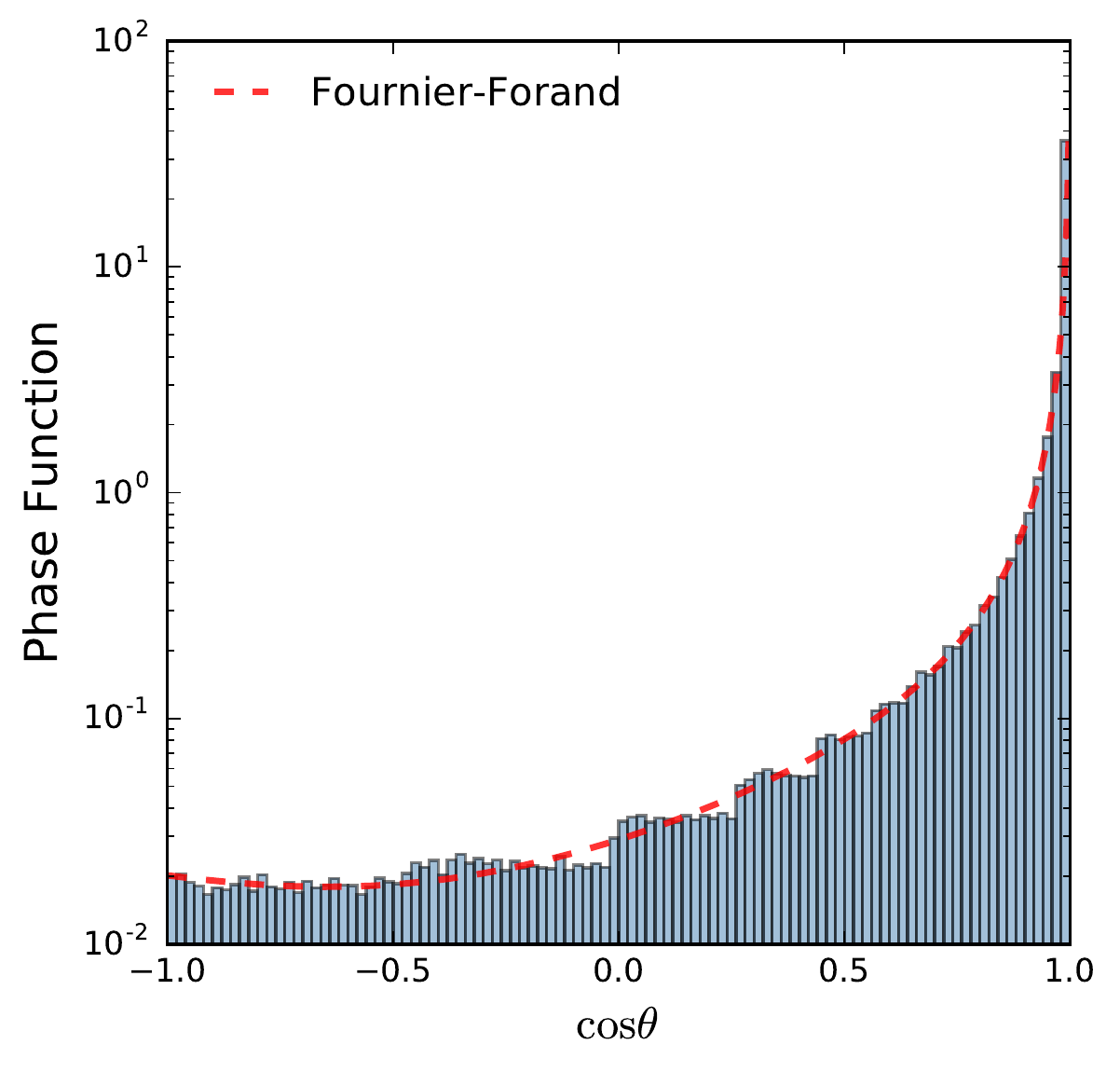}\label{ff:b}}
\caption{FF phase function (red dashed line) ccompared with the normalised histogram of 1000000 samples generated using the hybrid tabulated method with PCA (top panel) . The parameters are set as  $m = 1.1$ and $\alpha = 3.62$.}
\label{fig:ff}
\end{figure}
In the tabulated method, the range $[0, 1]$ of the CDF for the scattering phase function is divided uniformly into subintervals. Hence, the sampling space is also divided into subsets. First, we draw a uniform random number $\xi \sim \mathcal{U}(0, 1)$ first. Then, the subinterval where $\xi$ is located is determined, $\xi \in [F_{k}, F_{k+1})$ ($\{F_k\}$ represents values of the CDF.). A random number $\mu$ from the desired scattering phase function can be found by linear interpolation:
\begin{equation}
\mu = \frac{\mu_{k+1} - \mu_{k}}{F_{k+1} - F_{k}} (\xi - F_{k}) + \mu_k
\end{equation}
where $\mu_k$($-1=\mu_1 < \mu_2 < ... < \mu_{K+1} = 1$) represents the value corresponding to $F_{k}$. 
The index of the interval can be found using the bisection search algorithm\cite{Sedgewick}. It should be noted that in this widely used tabulated method, the scattering phase function is actually approximated by a piecewise constant function:
\begin{equation}
p(\mu) \approx \sum_{k = 1}^{K}c_k \mathbb{I}(\mu\in A_k)
\end{equation}
Here, $\mathbb{I}(\cdot)$ is the indicator function, $A_k$ represents the interval $(\mu_k, \mu_{k+1})$  and 
\begin{equation}
c_k = \left \{
\begin{array}{rl}
(F_{k+1} - F_{k})/ (\mu_{k+1} - \mu_{k}) &\text{if the CDF is known},\\
 (p(\mu_{k+1}) + p(\mu_{k}))/2s & \text{otherwise}
 \end{array}\right .
\end{equation}
with $s = \sum_k c_k (\mu_{k+1} - \mu_{k})$. As an example, Fig. \ref{fig:hgpca} illustrates a normalized histogram using samples generated with the tabulated method and piecewise constant approximation (PCA). However, the quality of random numbers generated from the tabulated method above depends on the number of subintervals. Moreover, PCA sometimes is not accurate enough to expand a regular phase function near the forward and backward scattering regions. The phase function can be better approximated by a piecewise linear function(PLA) that agrees well with the actual value at each of the points $-1=\mu_1 < \mu_2 < ... < \mu_{K+1} = 1$. Based on this observation, improvements can be made by extending PCA to PLA or the piecewise log-linear approximation (PLLA). For PLA, the corresponding approximate phase function can be represented as 
\begin{equation}
p(\mu) \approx \sum_{k = 1}^{K} (a_k \mu + b_k) \mathbb{I} (\mu\in A_k)
\end{equation}
where  $a_k = (p(\mu_{k+1}) - p(\mu_{k}))/ (\mu_{k+1} - \mu_{k})$ and $b_k $.The sampling procedure is similar to the one described above, except the following quadratic equation is solved:
\begin{equation}
\frac{1}{2}a_k \mu^2 + b_k \mu - \frac{1}{2}a_k \mu_k^2 - b_k \mu_k = \xi
\end{equation}
The procedure for sampling is shown in Algorithm \ref{alg:pla}. Fig. \ref{fig:hgpla} illustrates a normalised histogram of $\mu$ sampled from PLA of a HG phase function. And PLA outperforms PCA, as shown in Fig. \ref{fig:hgcdf} and Fig. \ref{fig:miecdf}. Further acceleration, as shown in Algorithm \ref{alg:acc}, can be made if we select the values of the CDF to be uniform on $[0, 1], $\cite{Zijp:94, Naglic:17}. By using such a setting, the index of subintervals can be found more easily. The trade-off would be intervals must be fine enough to approximate the scattering phase function near the backward scattering region.  

\begin{algorithm}
\caption{Tabulated method with PLA}\label{alg:pla}
\begin{algorithmic}[1]
\Procedure{Sample}{$\{\mu_i\}, \{F_i\}$}
\State $\xi \sim \mathcal{U}(0, 1)$
\State Find $k$ with $F_{k} \leq \xi \leq F_{k+1}$  \Comment{Bisection search}
\State Solve $\frac{1}{2}a_k \mu^2 + b_k \mu - \frac{1}{2}a_k \mu_k^2 - b_k \mu_k = \xi$
\State \textbf{return} $\mu\in [\mu_k, \mu_{k+1}]$
\EndProcedure
\end{algorithmic}
\end{algorithm}

Another improvement was established in this study by dividing the entire interval of $\mu \in [-1, 1]$ into equal probable subintervals similar to the tabulated method; however, for each subinterval, the accept-reject method is applied instead of interpolation. In particular, if the analytic form of the phase function is known, the phase function can be sampled accurately using this approach. The procedure for the hybrid tabulated method combined with the accept-reject method is shown in Algorithm \ref{alg:ntm}. Fig. \ref{fig:ff} compares the sample quality of the hybrid and the simple tabulated method with normalised histograms of random $\cos\theta$ generated from the FF phase function. As shown in Fig. \ref{fig:ff}, the FF phase function, which has long been considered impossible to exactly sample, can be sampled accurately using the hybrid tabulated method.
\begin{algorithm}
\caption{Tabulated method with PLA and acceleration}\label{alg:acc}
\begin{algorithmic}[1]
\Procedure{Sample}{$\{\mu_i\}, \{F_i\}, \{p_i\}$}
\State $\xi_1 \sim \mathcal{U}(0, 1)$
\State $k \gets \lfloor{\xi_1/(F_{2} - F_{1})}\rfloor$ 
\State $M \gets \max(p(\mu_k), p(\mu_{k+1}))$
\State \textbf{do} \Comment{Accept-reject sampling}
\State \qquad$\xi_2, \xi_3 \sim \mathcal{U}(0, 1)$
\State \qquad $p \gets (p_{k+1} - p_k)\xi_2  + p_k$ \Comment{Piecewise linear approximation}
\State \textbf{while} ($p < \xi_3 * M$)
\State \textbf{return} $\mu_k + \xi_2 (\mu_{k+1} - \mu_{k})$
\EndProcedure
\end{algorithmic}
\end{algorithm}

\begin{algorithm}
\caption{A hybrid sampling method by combing the accept-reject technique with interpolation..}\label{alg:ntm}
\begin{algorithmic}[1]
\Procedure{Sample}{$\{\mu_i\}, \{F_i\}$}
\State $\xi_1 \sim \mathcal{U}(0, 1)$
\State Find $k$ with $F_{k} \leq \xi_1 \leq F_{k+1}$  \Comment{Bisection search}
\State $M \gets \max(p(\mu_k), p(\mu_{k+1}))$
\State \textbf{do} \Comment{Accept-reject sampling}
\State \qquad$\xi_2, \xi_3 \sim \mathcal{U}(0, 1)$
\State \qquad$\mu \gets \mu_k + \xi_2 (\mu_{k+1} - \mu_{k})$
\State \textbf{while} ($p(\mu) < \xi_3 * M$)
\State \textbf{return} $\mu$
\EndProcedure
\end{algorithmic}
\end{algorithm}

\subsection{Gibbs sampling}
Tabulated methods are simple and easy to implement; however, the accuracy of the simulation depends on the number of interpolation points. Moreover, tabulated methods are computationally expensive, and they consume large amounts of memory to be implemented on GPUs. In the following section, Gibbs sampling, a sampling method using Markov chains is introduced for generating random numbers from certain complicated analytic scattering phase functions. 

The Gibbs sampling method relies on building a joint density with conditional distributions that are easy to simulate. Unlike other methods, auxiliary random variables are required in for Gibbs sampling. Random numbers are iteratively generated from a sequence of conditional distributions. The limit distribution of this sequence converges to the desired density

As an example, the Draine phase function is sampled using the Gibbs sampling method. The joint density required to be sample is
\begin{equation}
p(\mu, u|\alpha, g) \propto \mathbb{I}( u < 1 + \alpha \mu^2) \frac{1-g^2}{2(1+g^2-2g\mu)^{3/2}}
\end{equation}
where $u$ is an auxiliary random variable. The two conditional densities are 
\begin{eqnarray}
p( u|\mu,\alpha, g) &\propto& \mathbb{I}( u < 1 + \alpha \mu_{t-1}^2), \\
p(\mu|u, \alpha, g) &\propto& \frac{1-g^2}{2(1+g^2-2g\mu)^{3/2}},  \qquad 1+ \alpha \mu^2 < u_t.
\end{eqnarray}
The sampling procedure is summarised in Algorithm \ref{alg:draine}. 
\begin{algorithm}
\caption{Sample the Draine phase function}\label{alg:draine}
\begin{algorithmic}[1]
\Procedure{sample draine}{$ \mu_t $}
\State sample $\xi_1$ from $\mathcal{U}(0, 1+\alpha \mu_t^2)$\Comment{sample $Y\sim f_{Y|X}$}
\State \textbf{do}  \Comment{sample $X\sim f_{X|Y}$}
\State \qquad$\xi_2 \sim \mathcal{U}(0, 1)$
\State \qquad$\mu  \gets \frac{1}{2g}((1 + g^2) - ((1 - g^2)/(1 + g(2\xi_2 - 1)))^2)$
\State \textbf{while} ($ 1+\alpha \mu^2 < \xi_1$)
\State \textbf{return}  $\mu_{t+1}  \gets \mu$
\EndProcedure
\end{algorithmic}
\end{algorithm}

\begin{figure}
\centering
\includegraphics[width=0.8\linewidth]{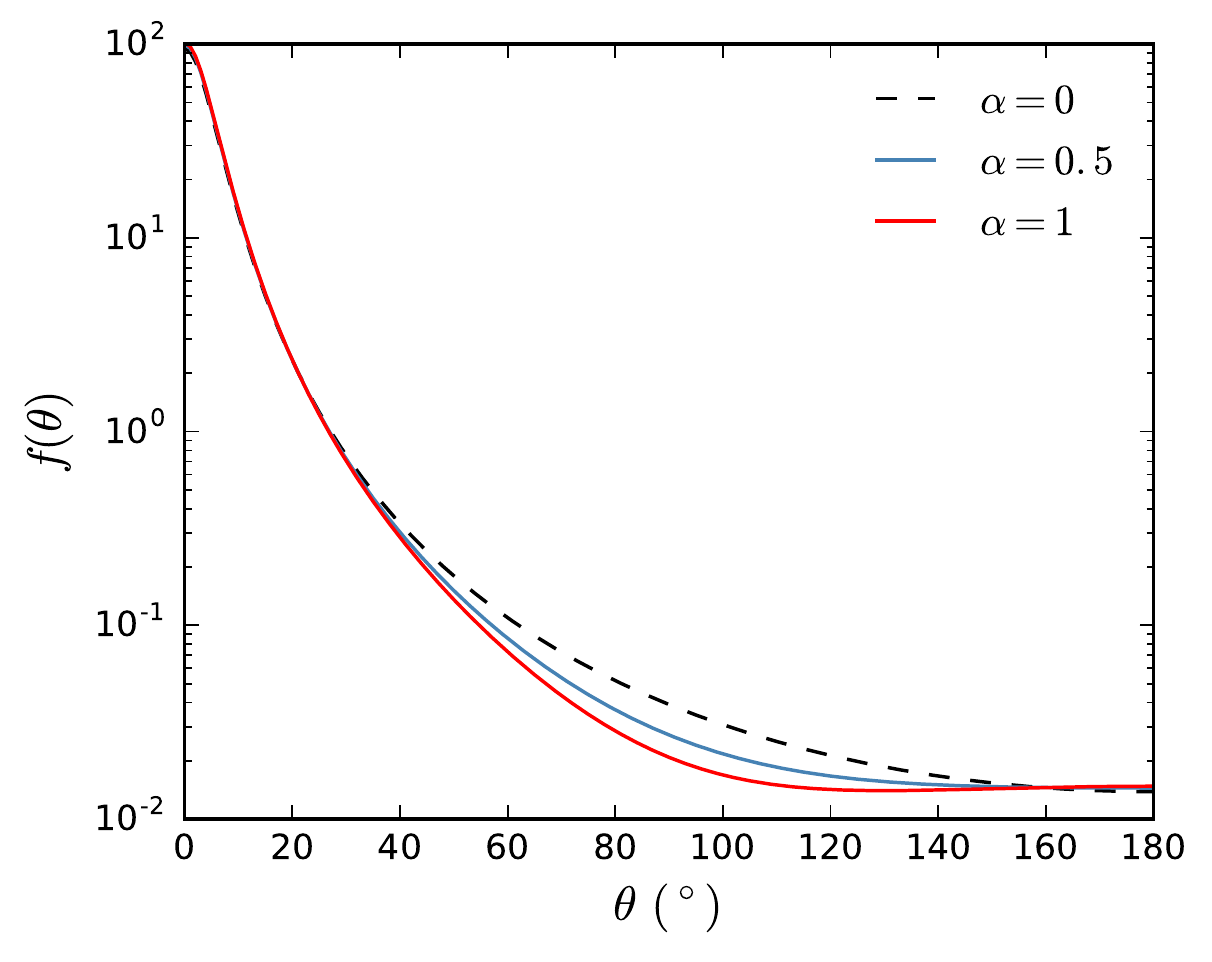}
\caption{Draine scattering phase function with different parameter values. It reduces to the HG phase function when $\alpha = 0$ and the Cornette-Shanks phase function when $\alpha = 1$.} 
\label{fig:drainephasefunction}
\end{figure}

\begin{figure}%
\centering
{\includegraphics[width=0.45\linewidth]{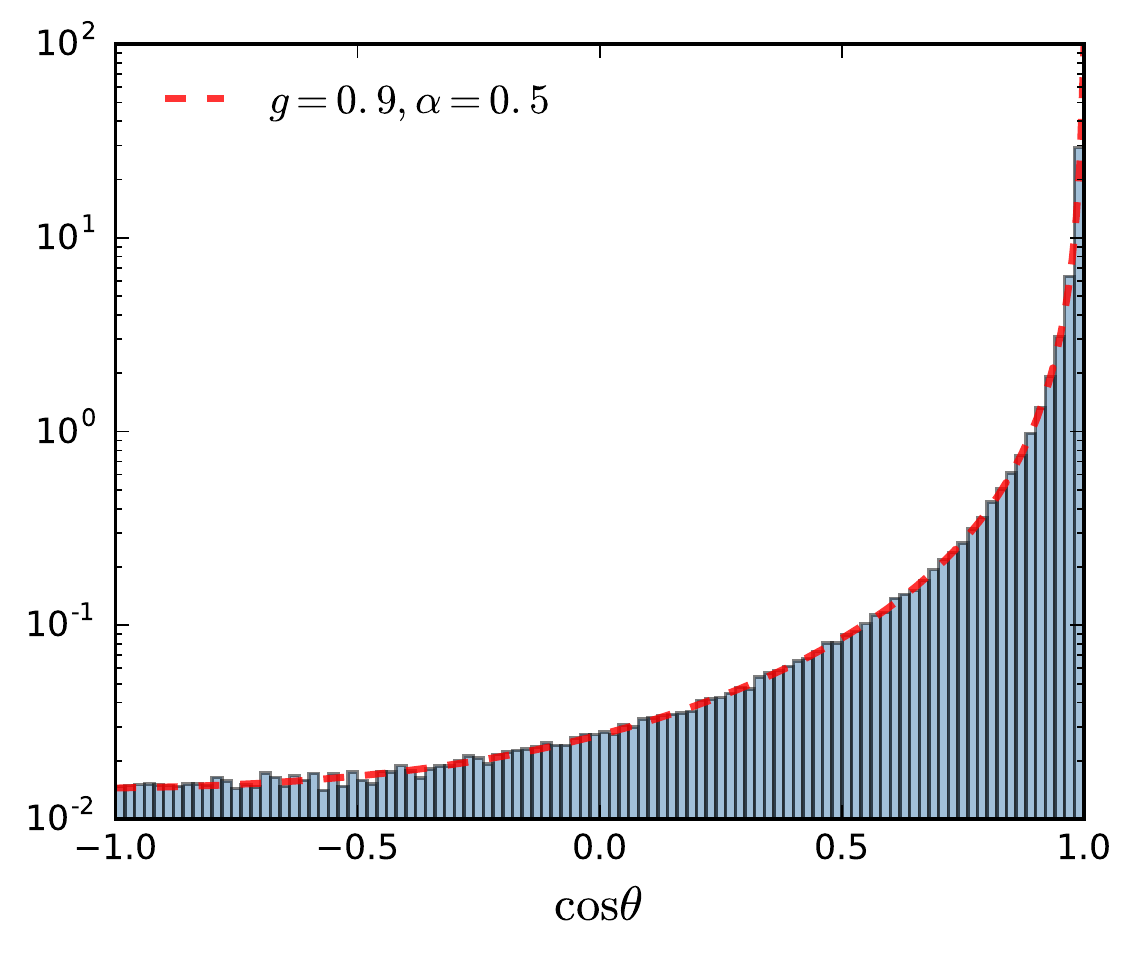} }%
\qquad
{\includegraphics[width=0.45\linewidth]{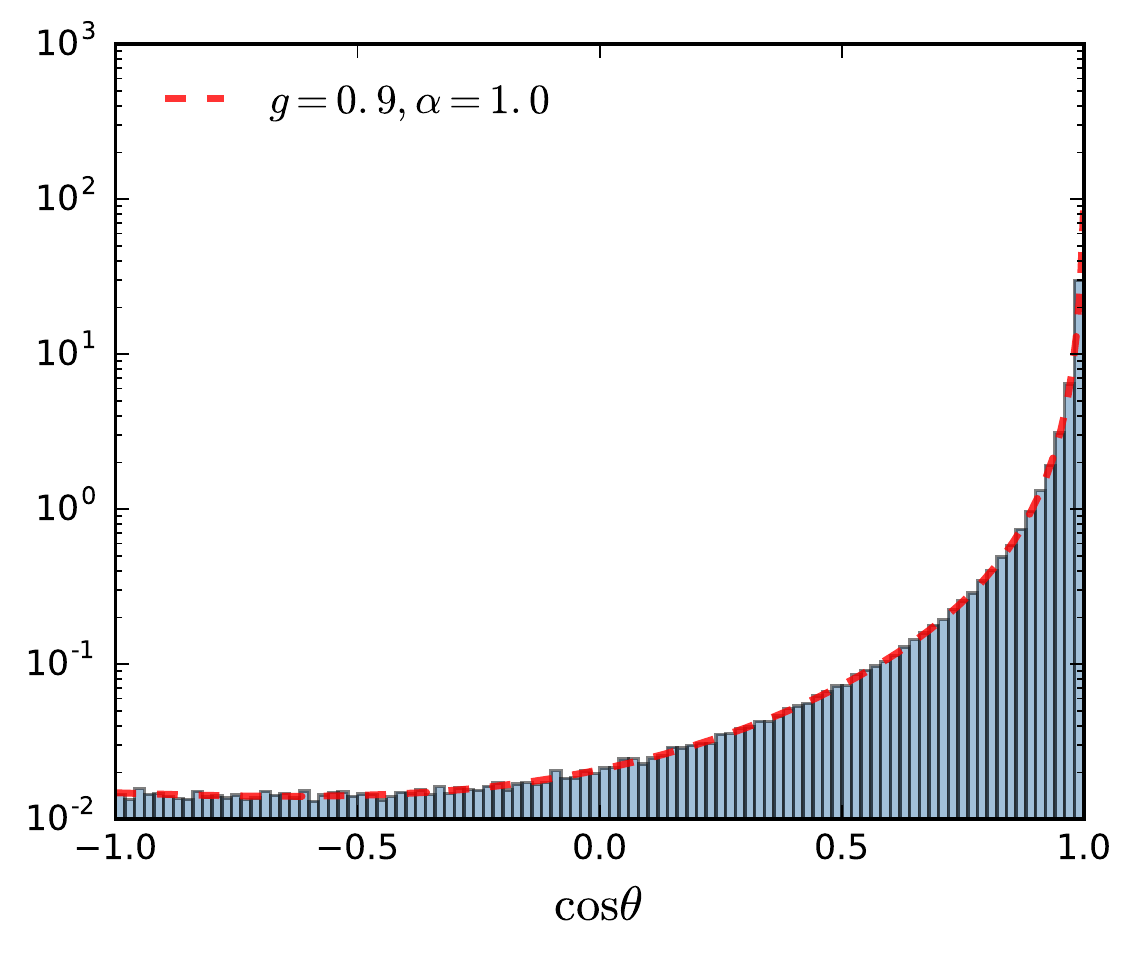} }%
\caption{Normalised sample histograms for the Draine scattering phase function compared with its analytic curve.}%
\label{fig:normedhistogramdraine}%
\end{figure}

\begin{figure}%
\centering
{\includegraphics[width=0.45\linewidth]{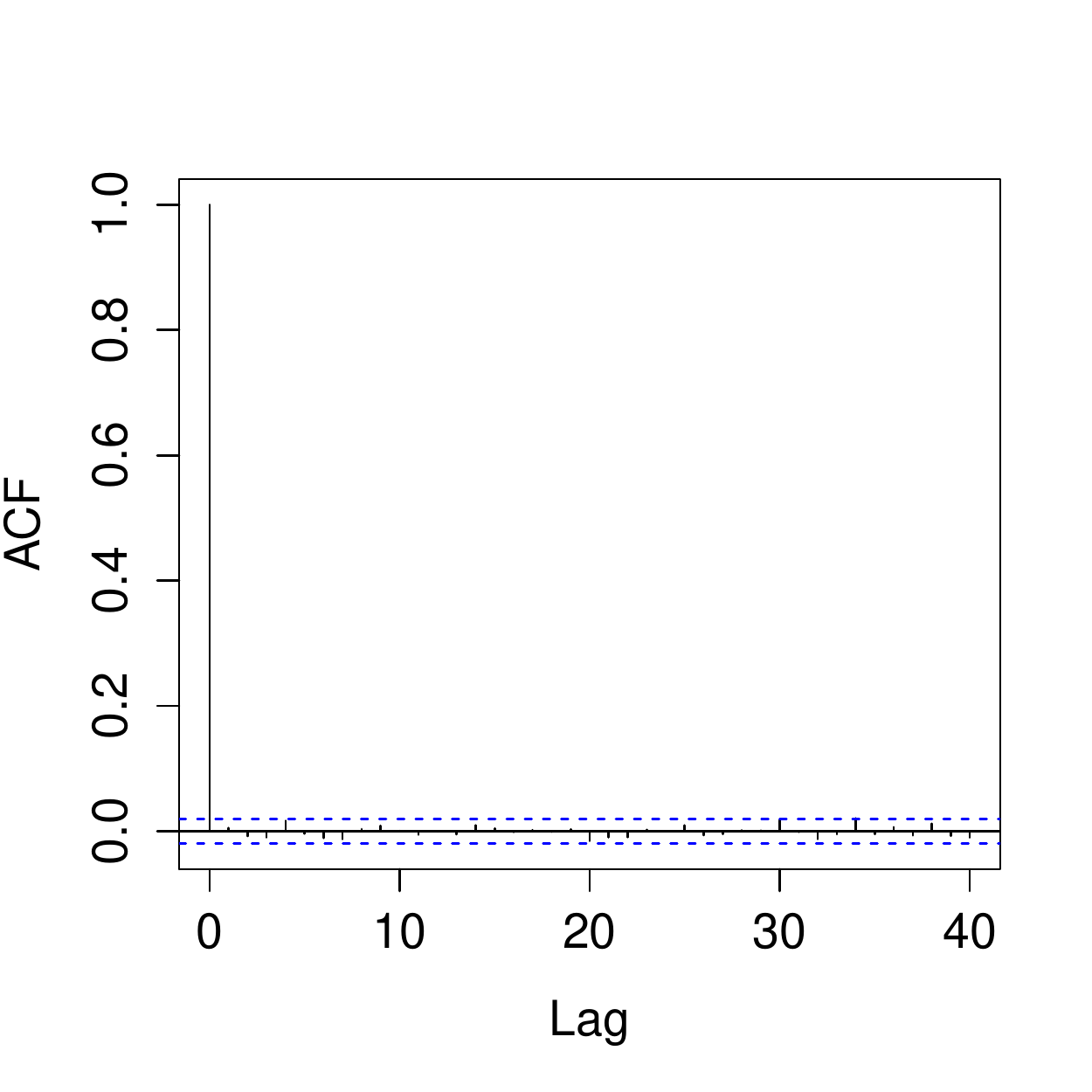} }%
\qquad
{\includegraphics[width=0.45\linewidth]{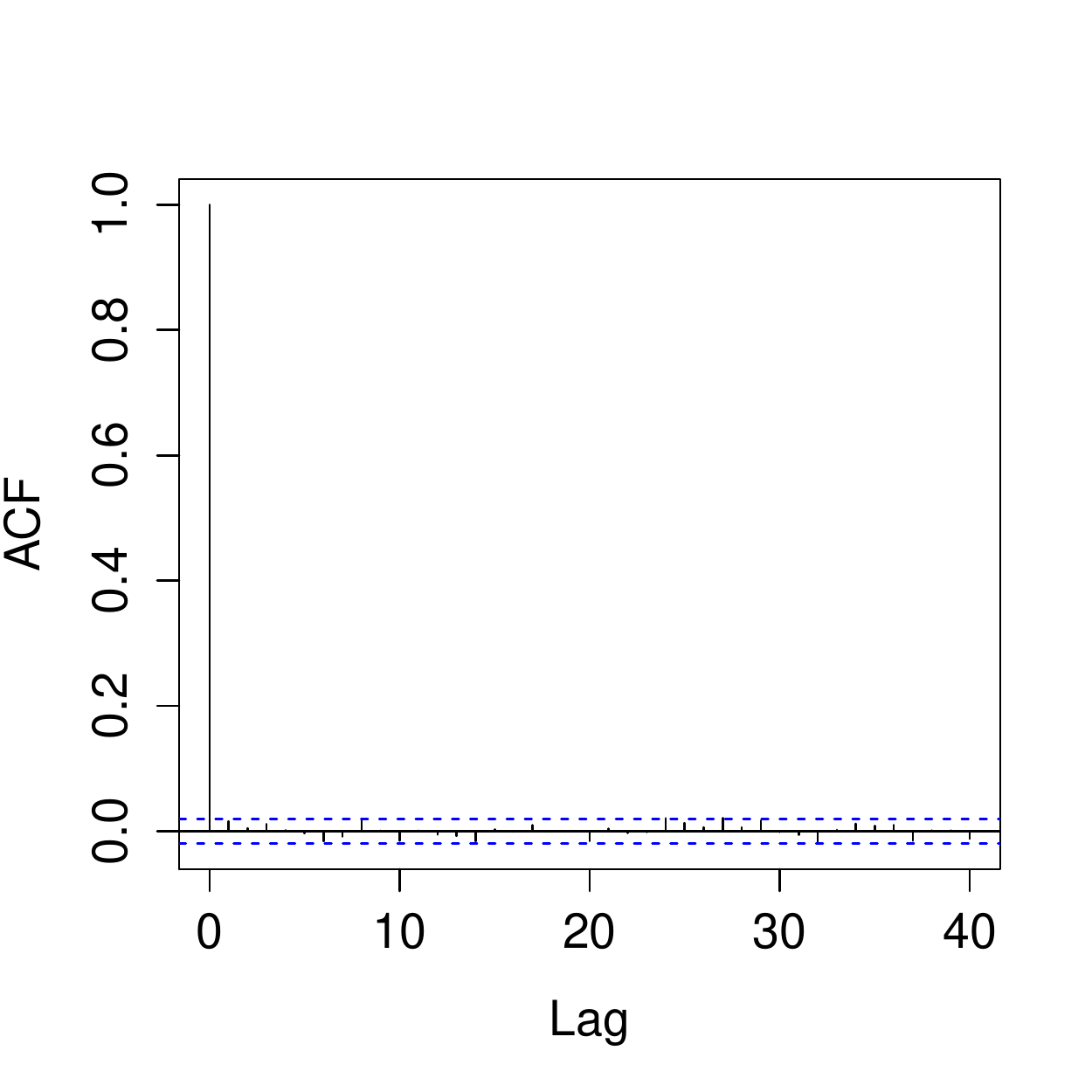} }%
\caption{Auto-correlation functions corresponding to samples from the Draine scattering phase function with (a) $g = 0.9$ and $\alpha = 0.5$, and (b) $g = 0.9$ and $\alpha = 1.0$.}%
\label{fig:acf1}%
\end{figure}

The Fig. ~\ref{fig:drainephasefunction} compares the Draine phase function with different parameters and Fig.~\ref{fig:normedhistogramdraine} shows normalised histograms for the Draine phase function samples generated using the Gibbs sampling method. The autocorrelation functions of random scattering angles are plotted in Fig.~\ref{fig:acf1}; from this figure, we can see that the autocorrelation of Lag-1 is considerably below 0.1. As expected, the Gibbs sampling method is an extremely powerful method that allows us to sample complicated analytical phase functions.

\section{Simulation and Results}
\begin{figure}%
\centering
{\includegraphics[width=0.65\linewidth]{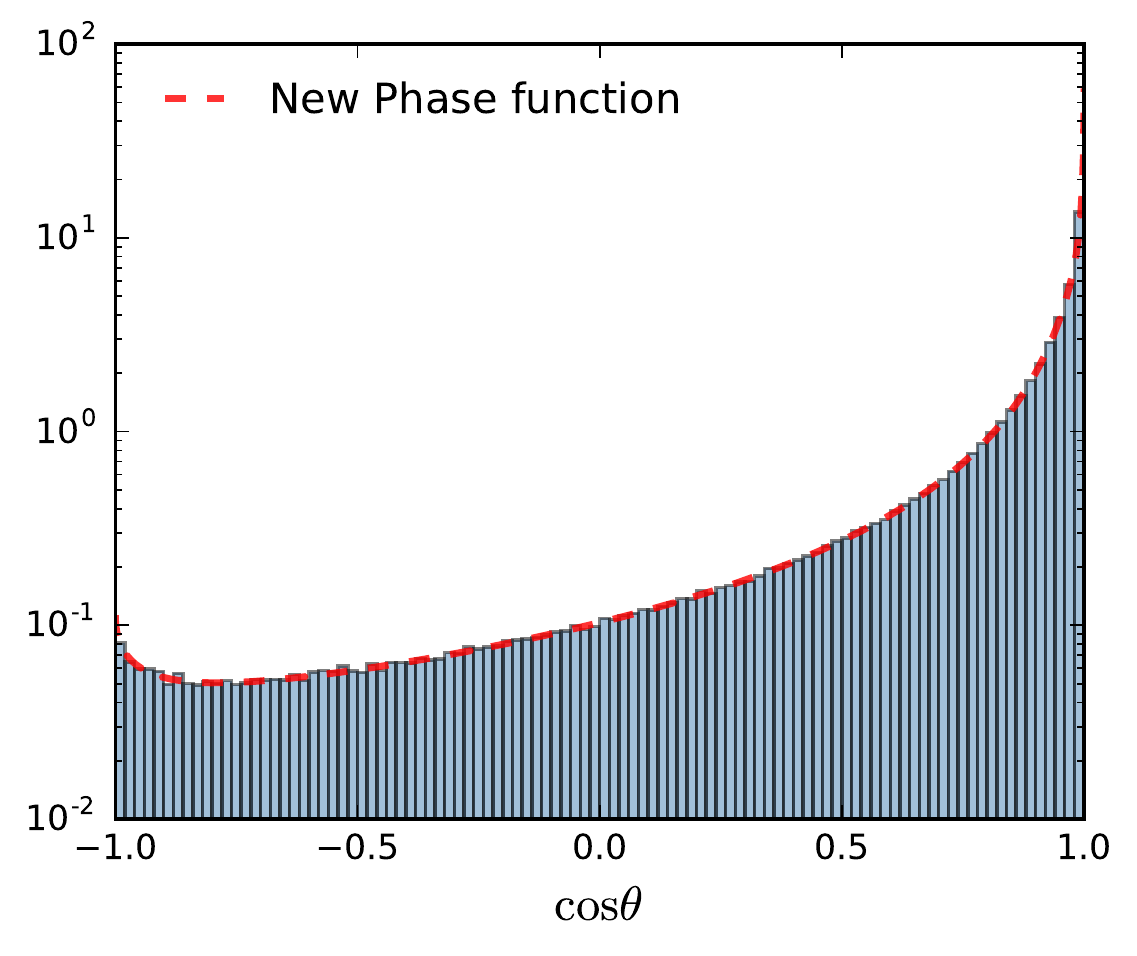} }%
\qquad
{\includegraphics[width=0.75\linewidth]{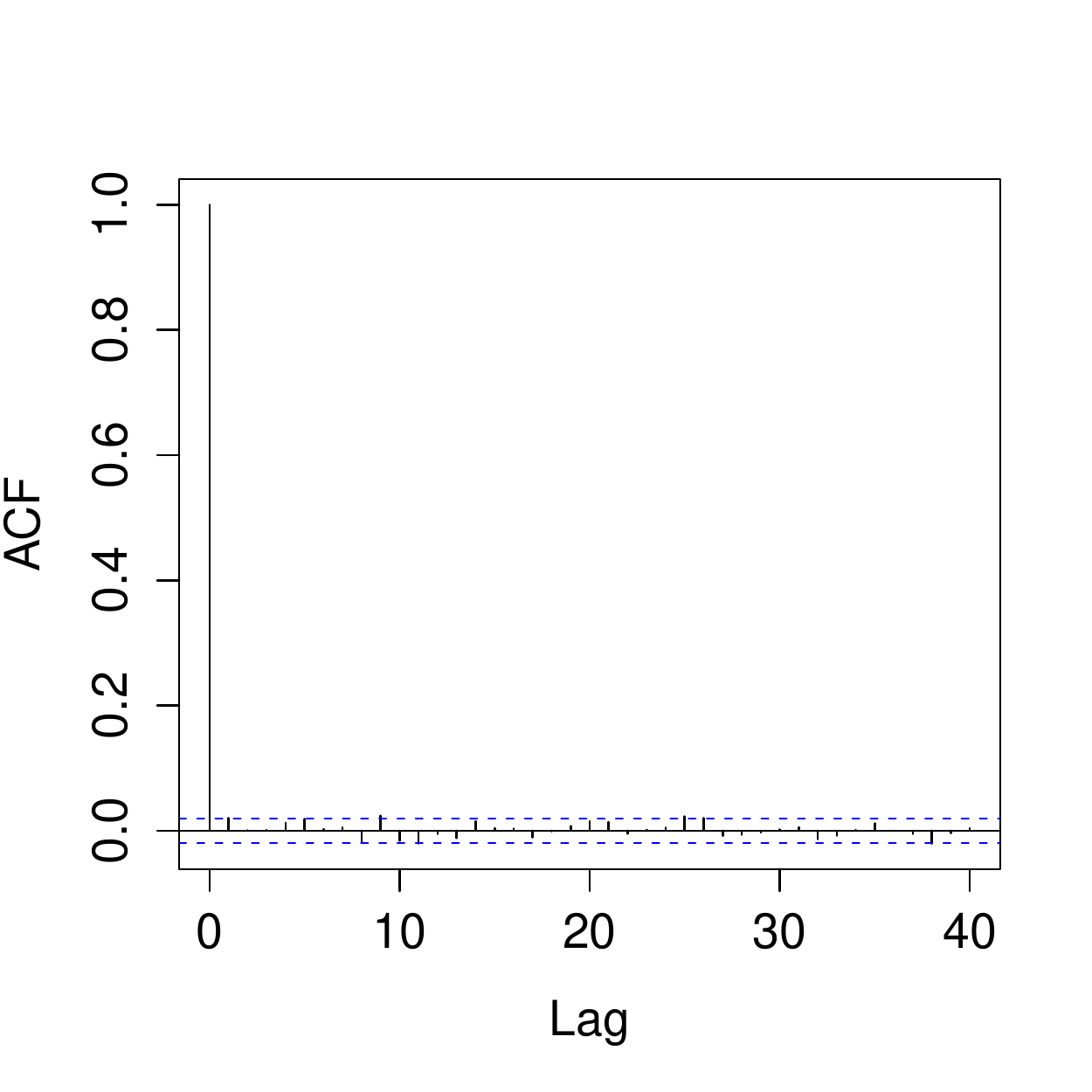} }%
\caption{(a) Normalised sample histograms for the new scattering phase function compared with its analytic curve. (b) Auto-correlation function corresponding to samples generated with Algorithm \ref{alg:mhg}.}%
\label{fig:newacf}%
\end{figure}

Analytic phase functions such as the Draine phase function can provide good approximations to realistic dust scattering at long wavelengths. For a higher frequency case, the true phase function of particles exhibits strong peaks (both forward and backward) and modelling using these analytic phase functions can result in noticeable errors. Considering two peaks of the true phase functions at the forward and backward directions, a new phase function is proposed with exponential decay terms:
\begin{eqnarray}
 f(\cos\theta | g, a, b, k, k') = C\frac{(1-g^2)(1 + a\exp(-k \theta)+b\exp(-k' (\pi-\theta) ) }{(1+g^2-2g\cos\theta)^{3/2}}
\end{eqnarray}
where $g$, $a$, $b$, $k$ and $k'$ are certain parameters, and $C$ is the normalisation constant that can be determined numerically.
\begin{figure}
\centering
\includegraphics[width=0.8\linewidth]{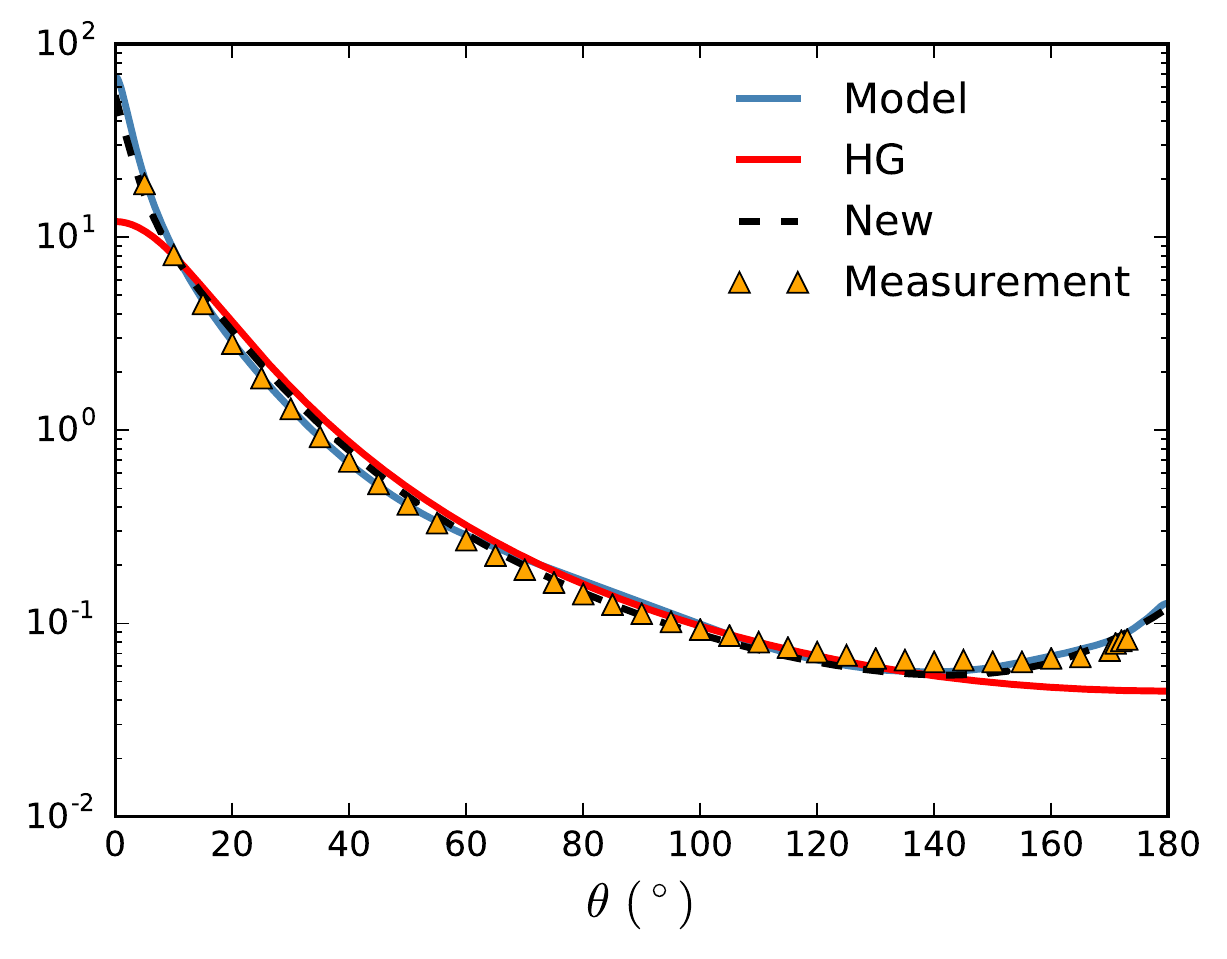}
\caption{ Scattering phase function and various approximations for feldspar particles. The measured phase function is obtained from the Amsterdam Light Scattering Database\cite{munoz:2012} (orange triangles) at a wavelength of 632.8 nm. The refractive index is 1.5 + 0.001i. The model averaged phase function is calculated with rough spheroids as described in \cite{ZHANG2016325}. }
\label{fig:spheroid}
\end{figure}
Fig.~\ref{fig:spheroid} compares the ensemble averaged phase function (modelled with a rough spheroid  \cite{ZHANG2016325} ) and the new phase function. 
The sampling process for the new phase function is similar, and the joint density is 
\begin{equation}
p(\mu, u| g, a, b, k, k') \propto \mathbb{I}(u < 1 + a\exp(-k \theta)+b\exp(-k' (\pi-\theta) ) ) \frac{1-g^2}{2(1+g^2-2 g\mu)^{3/2}}
\end{equation}
The sampling procedure is summarised in Algorithm \ref{alg:mhg}. Fig.~\ref{fig:newacf}a illustrates the normalised histogram of the sample from the new phase function using the Gibbs sampling method and the autocorrelation function plot of samples from this new phase function is shown in Fig.~\ref{fig:newacf}b. The figure shows no obvious short-term autocorrelation in the scattering angle sample, which indicates the independence of these samples.
\begin{algorithm}
\caption{Sample the new modified HG phase function}\label{alg:mhg}
\begin{algorithmic}[1]
\Procedure{sample npf}{$ \mu_i $}
\State $\theta_i \gets \arccos \mu_i$
\State sample $\xi_1 \sim \mathcal{U}(0, 1 + a\exp(-k \theta_i)+b\exp(-k' (\pi-\theta_i) ))$ \Comment{sample $Y\sim f_{Y|X}$}
\State \textbf{do}  \Comment{sample $X\sim f_{X|Y}$}
\State \qquad$\xi_2 \sim \mathcal{U}(0, 1)$
\State \qquad$\mu  \gets \frac{1}{2g}((1 + g^2) - ((1 - g^2)/(1 + g(2\xi_2 - 1)))^2)$
\State \qquad$\theta \gets \arccos \mu$
\State \textbf{while} ($ 1 + a\exp(-k \theta)+b\exp(-k' (\pi-\theta) ) < \xi_1$)
\State \textbf{return}  $\mu_{i+1}  \gets \mu$
\EndProcedure
\end{algorithmic}
\end{algorithm}

As a more realistic test, a backward Monte Carlo \cite{manfred} simulation is conducted to compute the reflected and transmitted radiance through a plane-parallel medium. Each photon is traced from the detector to a light source in a backward manner. The radiance is calculated by summing the contribution of each photon. The albedo of the bottom boundary is set to 0, and only volume scattering events are considered here. The new phase function is used to model dust scattering and simulated using the Gibbs sampling method. The measured scattering phase function obtained from the Amsterdam Light Scattering Database is used as the true dust scattering phase function, sampled using Algorithm \ref{alg:pla}.

\begin{figure}
\centering
{\includegraphics[width=0.45\linewidth]{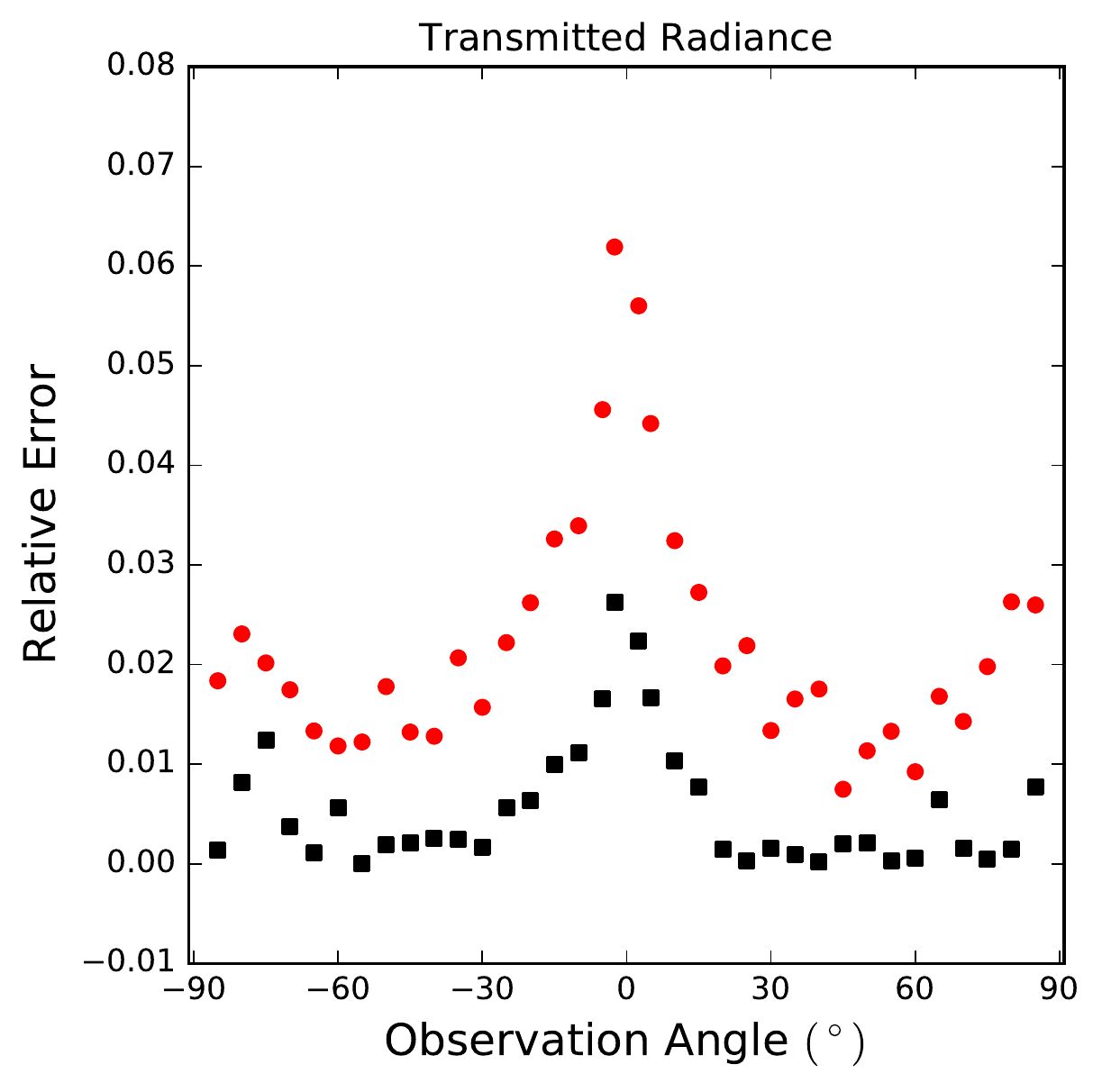} }%
\qquad
{\includegraphics[width=0.45\linewidth]{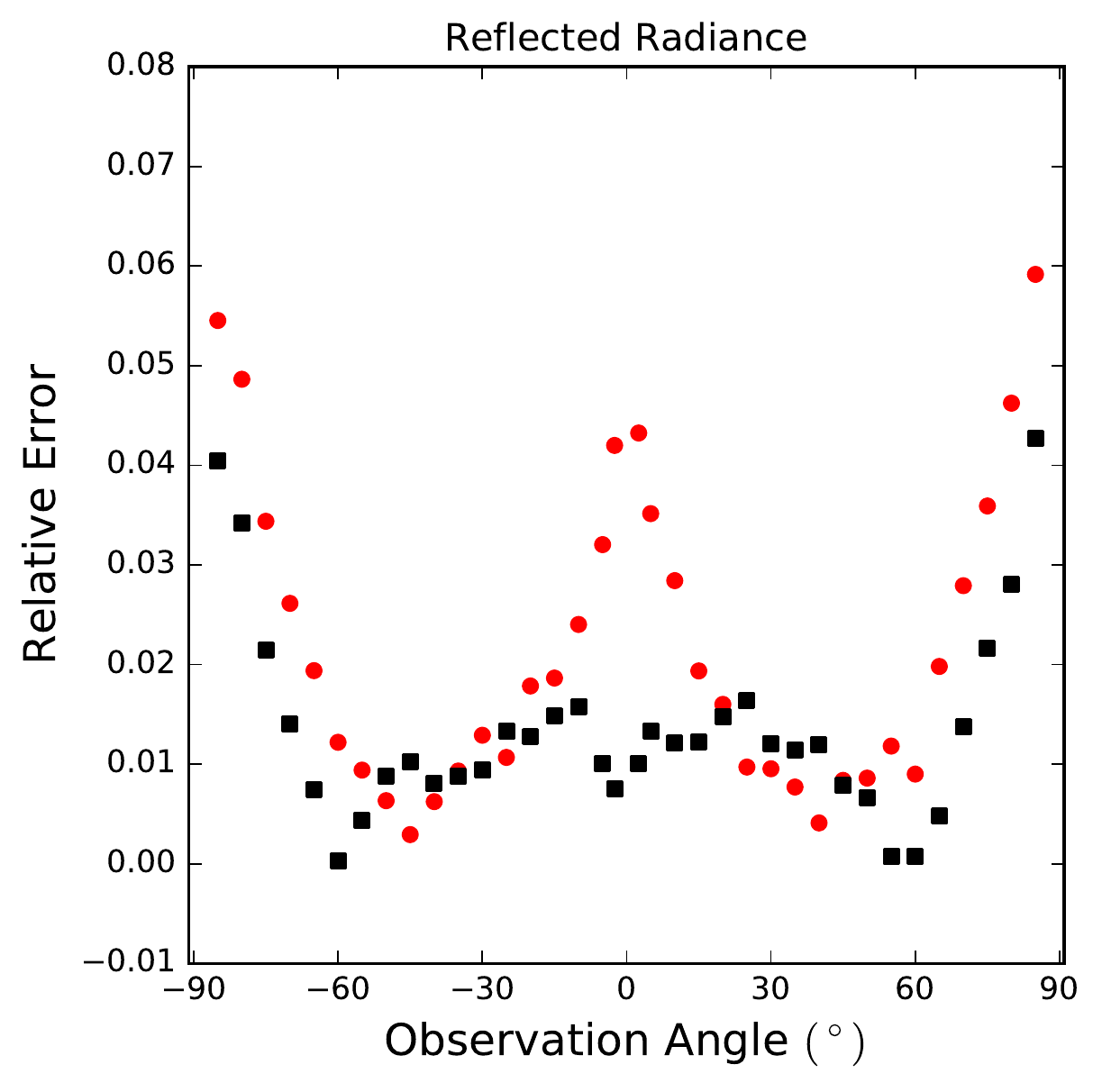} }%
\caption{Relative error as a function of observation zenith angle for simulated transmitted and reflected radiance of a turbid medium. A positive angle means that the azimuth angle is $0^{\circ}$, while a negative angle means that the azimuth angle is $180^{\circ}$.The red circles represent the relative error for results simulated with the HG phase function, and the black squares represent the relative error for results simulated with the new phase function. The exact phase function was sampled with the tabulated method. The new phase function was sampled using the Gibbs sampling method. The HG phase function was sampled with the inverse CDF method. }%
\label{fig:TRRR}%
\end{figure}

The left panel of Fig.~\ref{fig:TRRR} illustrates the relative error of the simulated transmitted radiance with the HG phase function and the new phase function. The error with the new phase function is below 2\% for nearly all angles and approximately $1/2$ of the error with the HG phase function. The right panel of Fig.~\ref{fig:TRRR} illustrates the relative error for the simulated reflected radiance with these two phase functions. The result simulated with the new phase function also outperforms that with the HG phase function, except for only a small range of angles. For observation angles from $-30^{\circ}$ to $30^{\circ}$, the relative error corresponding to the new phase function is roughly constant with a value of approximately 1\%. Conversely, the relative error corresponding to the HG phase function reaches a maximum at the backward direction and is approximately five times larger than the relative error corresponding to the new phase function. Consequently, results simulated with the new phase function fit the realistic dust scattering results nearly uniformly better than those simulated with the widely used HG phase function. 

\section{Discussion and Conclusion}

In this paper, several new sampling methods were proposed for simulating the scattering phase function in the MCRT algorithm. We show that the commonly used tabulated method actually samples a PCA of the exact scattering phase function, which may not be a good choice for modelling realistic dust scattering. Improvements are made by extending the PCA to a PLA or a PLLA. This modification requires minor revisions in the code of the MCRT algorithm and is not computationally expensive. Furthermore, a hybrid method combining the tabulated method and the accept-reject method is proposed by substituting the interpolation procedure with an accept-reject procedure. As an example, the regularized FF function is sampled exactly with this hybrid method. In addition, the Gibbs sampling method is also recommended for generating random numbers from certain complicated analytic phase functions; this has previously not been used in MCRT. Based on the HG phase function, a new phase function with two exponential decay terms is also proposed to better model realistic dust scattering. We also demonstrate that this new phase function can be sampled efficiently using the Gibbs sampling method. The performance of this new phase function is validated for a plane-parallel medium simulated using the backward MCRT algorithm. The results simulated with the new phase function fit the realistic dust scattering model considerably better than those simulated with the HG phase function, particularly near the backscattering region. It is recommended that this new phase can be used for modelling radiation transport in dusty planetary atmospheres and interstellar mediums. 

\section*{Acknowledgments}
We would like to thank anonymous reviewers and the review editor for their helpful comments. Part of this work was supported by the National Natural Science Foundation of China (NSFC) (Grant No. 41705010) and the Fundamental Research Funds for the Central Universities (DUT19RC(4)039).

\bibliographystyle{unsrt}  
\bibliography{references}  


\end{document}